\documentclass[usenatbib,usegraphicx,useAMS]{mn2e}
\usepackage{amsmath}
\usepackage{amssymb}

\title{Hypervelocity Collisions of Binary Stars at the Galactic Centre}

\author[Idan Ginsburg \& Abraham Loeb]{Idan Ginsburg\thanks{E-mail:
iginsburg@cfa.harvard.edu} \& Abraham
Loeb\thanks{E-mail:aloeb@cfa.harvard.edu}\\ Harvard-Smithsonian Center for
Astrophysics, 60 Garden St., MS 51, Cambridge, MA 02138, USA\\}

\begin{document}
\maketitle

\begin{abstract}

Recent surveys have identified seven hypervelocity stars (HVSs) in the halo
of the Milky Way.  Most of these stars may have originated from the breakup
of binary star systems by the nuclear black hole SgrA*.  In some instances,
the breakup of the binary may lead to a collision between its member
stars. We examine the dynamical properties of these collisions by
simulating thousands of different binary orbits around SgrA* with a direct
N-body integration code.  For some orbital parameters, the two stars
collide with an impact velocity lower than their escape velocity and may
therefore coalesce.  It is possible for a coalescing binary to have
sufficient velocity to escape the galaxy. Furthermore, some of the massive
S-stars near Sgr A* might be the merger remnants of binary systems, however
this production method can not account for most of the S-stars.

\end{abstract}

\begin{keywords}
black hole physics-Galaxy:center-Galaxy:kinematics and dynamics-stellar
dynamics
\end{keywords}

\section{Introduction} \label{I}

First theorized by \citet{Hills:88}, a hypervelocity star (HVS) has
sufficient velocity to escape the gravitational pull of the Milky Way
galaxy.  The first HVS, SDSS J090745.0+024507, was recently discovered in
the Galactic halo \citep{Brown:05,Fuentes:05}.  This HVS is located at a
heliocentric distance of $\sim 110$ kpc and has radial velocity 853 $\pm$ 12
kms$^{-1}$, over twice that needed to escape the gravitational pull of the
Milky Way. Since that initial discovery, six other HVSs have been
identified (\citealt{Edelmann:05}; \citealt{Hirsch:05};
\citealt{Brown:06a}; \citealt{Brown:06b}). \citet{Hills:88} suggested that
a HVS might result from a close encounter between a tightly bound binary
star system and the black hole at the Galactic center,
SgrA*. \citet{Yu-Tremaine:03} refined Hills' argument and noted that HVSs
might also be produced by three-body interactions between a star and a
binary black hole system.  Because the existence of a second
(intermediate-mass) black hole in the Galactic center \citep{Hansen:03} is
only a hypothetical possibility \citep{Scho:03}, we focus here on the
disruption of a tightly bound binary by a single supermassive black hole
(SMBH) with a mass of $\sim 4 \times 10^6 M_{\odot}$ whose existence is
robustly supported by data (e.g. \citealt{Ghez:05};
\citealt{Reid-Brunthaler:04}; \citealt{Scho:03}).

Simulations show that tight binaries can produce HVSs with velocities
comparable to the observed HVSs (e.g. \citealt{Ginsburg:06} (hereafter
Paper I); \citealt{Bromley:06}).  In Paper I, we show that the companion to
the hypervelocity star will be left in a highly eccentric orbit, which
agrees with the known orbits of a number of S-stars orbiting Sgr A* 
(e.g. \citealt{Eckart-Genzel:97}, \citealt{Scho:03}, and \citealt{Ghez:05}).
Therefore, we suggested that some of these stars are former
companions of HVSs.  Furthermore, a small fraction ($\sim 10\%$) of the
binary systems were found to collide. Here we examine in detail the
dynamical properties of such collisions, and check whether some of these
collisions may end in coalescence.

In \S 2 we describe the N-body code and the simulation parameters that were
used.  In $\S3$ we discuss our numerical results for the collisions, and in
$\S4$ we discuss the outcome of binary mergers at the Galactic Centre. Our
goal is not to cover the entire range of binaries that could produce HVSs
or end in collisions, but rather to determine whether some tight binaries
with masses similar to the HVSs observed thus far could coalesce.

\section{Computational Method} \label{CM}

In our study we have used the N-body code written by Aarseth
\citep{Aarseth:99} whose details were described in Paper I.  We treat the
stars as point particles and ignore tidal and general relativistic effects
on their orbits, since these effects are small at the distance ($\sim
10$AU) where the binary is tidally disrupted by the SMBH.  We have set the
mass of the SMBH to $M=4\times 10^6M_{\odot}$. The masses of the binary
members are set to either $3M_{\odot}$ \& $3M_{\odot}$ [since $3M_{\odot}$
is the estimated mass of SDSS J090745.0+024507
\citep{Fuentes:05}], or to $3M_{\odot}$ \& $10M_{\odot}$ [as
$10M_{\odot}$ is comparable to the estimated mass of HE 0437-5439
\citep{Edelmann:05}]. All runs start with the center of the circular binary
located 2000 AU ($=10^{-2}$pc) away from the SMBH along the positive
$y$-axis. This distance is comparable to the inner scale of the observed
distribution of stars around SgrA* (\citealt{Eckart-Genzel:97};
\citealt{Scho:03}; \citealt{Ghez:05}), allowing the remaining star to
populate this region after the ejection of its companion.  This radius is
also much larger than the binary size or the distance of closest approach
necessary to obtain the relevant ejection velocity of HVSs, making the
simulated orbits nearly parabolic.

We ran two sets of data.  The first had the binary system rotating along
the $x$--$y$ plane and the second along the $y$--$z$ plane. We used the
same initial distance for all runs to make the comparison among them easier
to interpret as we varied the distance of closest approach to the SMBH or
the relative positions of the two stars within the binary. We chose initial
binary separations between $a=0.05$ and $0.2$AU because such a range is
likely to produce HVSs for the above parameters (see Paper I).
Significantly wider binaries would give lower ejection velocities
\citep{Gualandris:05}.  Much tighter binaries would not be easily disrupted
by the black hole, or may coalesce to make a single star before interacting
with the SMBH. The radius of a main sequence star of a few solar masses is
$\sim 0.01$AU, and that of a 10 solar mass star is $\sim 0.03$AU (see,
e.g. Fig. 4 in Freitag \& Benz 2005).
Binaries tighter than $\sim 0.02$AU are precluded because the two stars
will develop a common envelope and eventually coalesce.

In the Galactic disk, about one-third to half of all stars form in binaries or small
multiple systems (see e.g. \citealt{Lada:06}; \citealt{Duquennoy-Mayor:91}), 
with roughly
equal probability per logarithmic interval of separations,
$dP/d\ln(a)=const$ (e.g. \citealt{Abt:83}; \citealt{Heacox:98};
\citealt{Larson:03}). In the Galactic center environment, the maximum
binary separation is limited by the tidal force of SgrA* at the distance
$d$ where the binary is formed (for conditions that enable star formation
near the SMBH, see \citealt{M-Loeb:04}).  Since the mass of the black hole
is $\sim 10^6$ times larger than that of a star, this implies a maximum
binary separation less than $(10^{-6})^{1/3}=10^{-2}$ of the initial
distance $d$. For $d=2\times 10^3$AU, the upper limit on the binary
separation would be 20AU (or smaller if the tidal restriction applies
during the formation process of the binary). If we assume a constant
probability per $\ln(a)$ for $0.02<a<20$AU, then the probability of finding
a binary in the range of $a=0.05$--$0.2$AU is substantial, $\sim 20\%$.

As shown in Paper I, the initial phase of the binary orbit plays a crucial
role in the outcome. Therefore, we sampled cases with initial phase values
of 0-360 degrees in increments of 15$^\circ$.  As initial conditions,
we gave the binary system no radial velocity but a tangential velocity with
an amplitude in the range between $5$ and $25~{\rm km~s^{-1}}$ at the
distance of 2000AU. In total, we ran 2000 simulations.

\section{Properties of Hypervelocity Collisions} \label{LC}

Given a binary system with stars of equal mass $m$ separated by a distance
$a$ and a SMBH of mass $M\gg m$ at a distance $b$ from the
binary, tidal disruption would occur if $b\la b_{\rm t}$ where
\begin{equation}
\frac{m}{a^3} \sim \frac{M}{{b^3_{\rm t}}}
\end{equation}
The distance of closest approach in the initial plunge of the binary
towards the SMBH can be obtained by angular momentum conservation from its
initial transverse speed $v_{\perp}$ at its initial distance from the SMBH,
$d$,
\begin{equation}
v_{\perp}d = \left(\frac{GM}{b}\right)^{1/2}b .
\end{equation}
The binary will be tidally disrupted if its initial transverse speed
is lower than some critical value,
\begin{equation}
v_\perp\la v_{\perp,\rm crit} \equiv {(GMa)^{1/2}\over d}\left({M\over
m}\right)^{1/6}= 10^2 {a_{-1}^{1/2} \over m_{0.5}^{1/6} d_{3.3}}
~{\rm {km~s^{-1}}},
\label{eq:crit}
\end{equation}
where $a_{-1}\equiv ({a}/{0.1~{\rm AU}})$, $d_{3.3}=(d/2000~{\rm AU})$,
$m_{0.5} \equiv (m/3M_{\odot})$. If $v_\perp\la v_{\perp,\rm crit}$, one
of the stars receives sufficient kinetic energy to become unbound, while
the second star is kicked into a tighter orbit around the SMBH.  The
ejection speed $v_{\rm ej}$ of the unbound star can be obtained by
considering the change in its kinetic energy $\sim v\delta v$ as it
acquires a velocity shift of order the binary orbital speed $\delta v \sim
\sqrt{Gm/a}$ during the disruption process of the binary at a distance
$\sim b_t$ from the SMBH when the binary center-of-mass speed is $v\sim
\sqrt{GM/b_t}$ \citep{Hills:88,Yu-Tremaine:03}. At later times, the binary
stars separate and move independently relative to the SMBH, each with its
own orbital energy.  For $v\la v_{\perp,\rm crit}$, we therefore expect
\begin{align}
v_{\rm ej} \sim \left[\left({\frac{Gm}{a}}\right)^{1/2}\left(
{\frac{GM}{b_{\rm t}}}\right)^{1/2}\right]^{1/2} \nonumber\\ = 
1.7 \times 10^3 
m^{1/3}_{0.5}a^{-1/2}_{-1} ~{\rm km~s^{-1}}.
\label{eq:model}
\end{align}

\begin{table}
\begin{tabular}{|r|r|r|r|r|r|}
\hline
$a$ (AU)&P(3$M_{\odot})$&P(10$M_{\odot}$)\\
\hline
0.05&0.11${\pm}$0.02&0.21${\pm}$0.05& & \\
0.10&0.11${\pm}$0.02&0.13${\pm}$0.04& & \\
0.15&0.06${\pm}$0.01&0.12${\pm}$0.03& & \\
0.20&0.03${\pm}$0.01&0.04${\pm}$0.02& & \\
\hline
0.05&0.09&0.27\\
0.10&0.06&0.13\\
0.15&0.03&0.09\\
0.20&0.02&0.07\\
\hline
\end{tabular}
\caption{Collision probability with different values of $a$ for binaries of
3M$_{\odot}$ \& 3M$_{\odot}$ (second column) and 10M$_{\odot}$ \&
3M$_{\odot}$ (third column).  The top four rows show the values obtained
from our simulations with their corresponding Poisson errors.  For
comparison, the bottom rows show the expected probability from a simplistic
{\it ``billiard ball''} model (without gravitational focusing) in which the
probability of a collision is ${{2(R_1+R_2)}/{2{\pi}a}}$. Here
$\{R_i\}_{i=1,2}$ are the radii of the two stars and $a$ the binary
separation. }
\label{tab}
\end{table}

\begin{figure*}
\begin{center}
\begin{tabular}{ccc}
\includegraphics[width=0.52\textwidth]{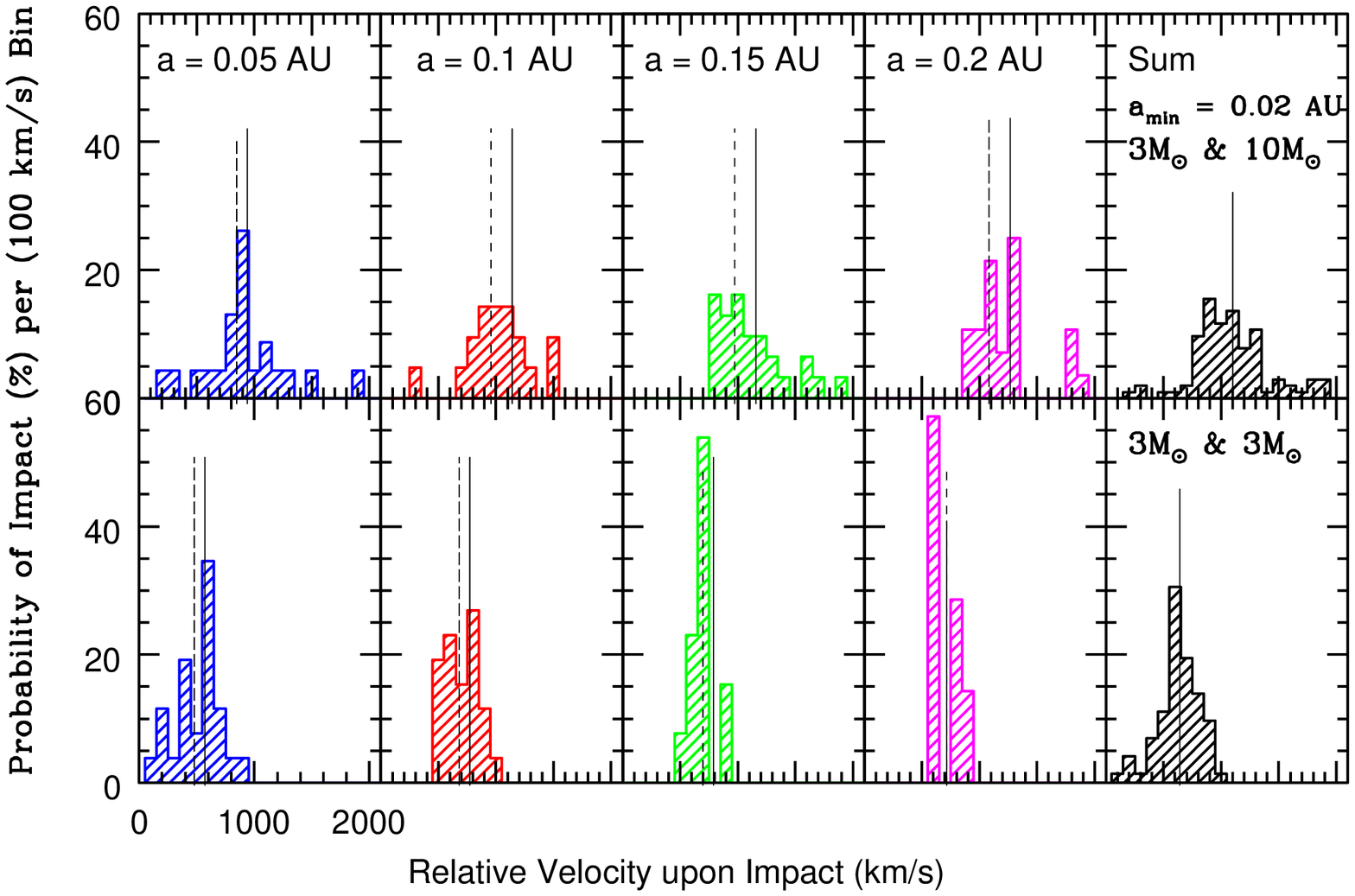}
&
\includegraphics[width=0.52\textwidth]{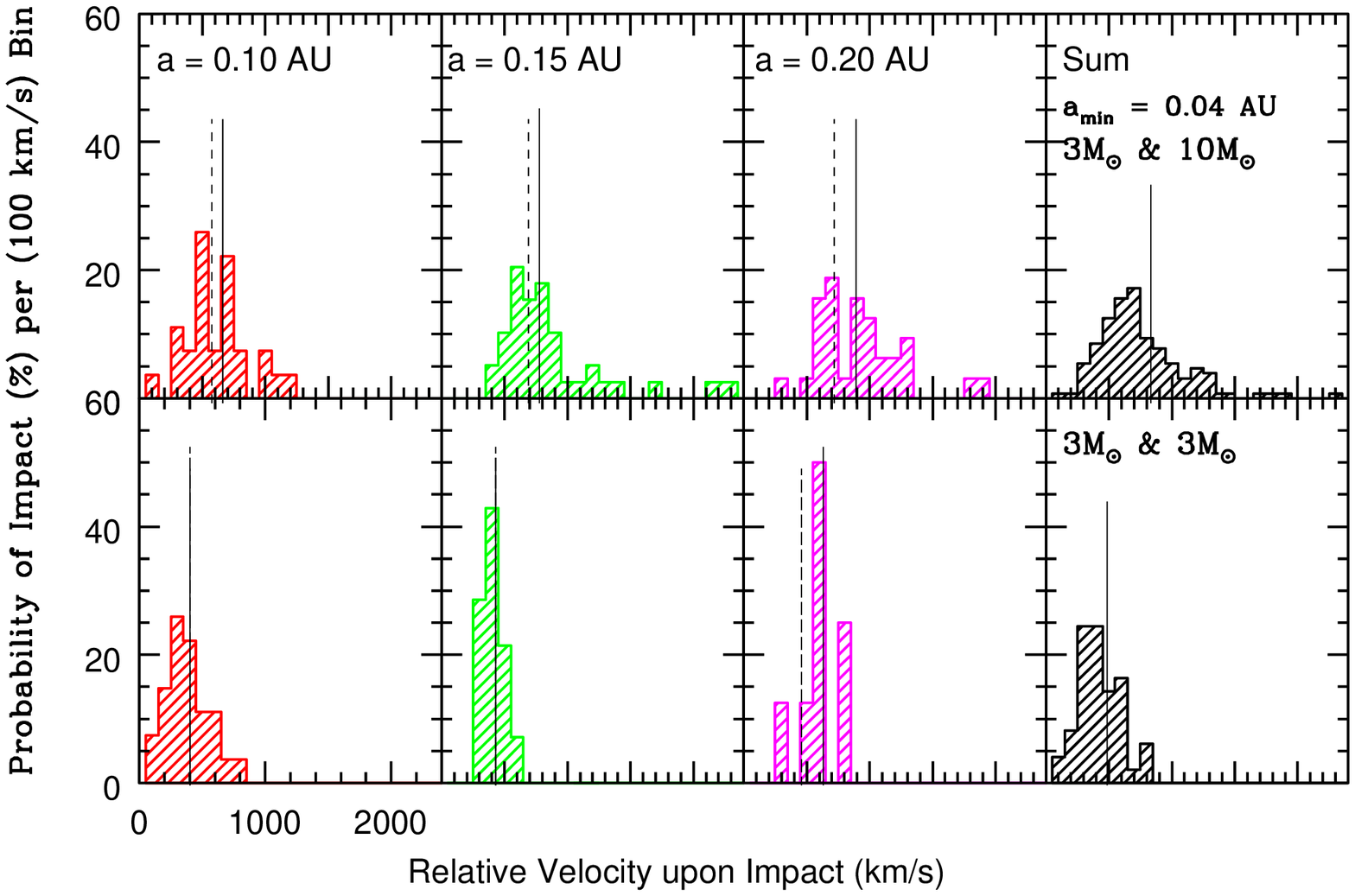}
&
\\
\end{tabular}
\end{center}
\caption{Fraction of all collisions (in percent per 100 km$s^{-1}$ bin)
versus relative velocity upon impact (in ${\rm km~s}^{-1}$). The left
section is for $a_{min}$ = 0.02 AU and the right section is for $a_{min}$ =
0.04 AU.  The label of the lower left panel corresponds to all panels.
The dashed vertical line shows the impact velocity that would
have resulted from free fall starting at the binary separation (see equation
\ref{eq:velocity}).  The solid line is the median velocity of all
runs. (We choose to use the median rather than the average value
because outliers bias the data otherwise.)
The minimum impact parameter
for a collision is expected to be $a_{min} = (R_1 + R_2)$ = 0.02 AU for a
3M$_{\odot}$ \& 3M$_{\odot}$ binary, but $a_{min}$ = 0.04 AU for a
10M$_{\odot}$ \& 3M$_{\odot}$ binary. We show results in other cases for
pedagogical purposes, namely to illustrate the dependence of the results on
the binary masses and $a_{min}$ separately. If the 10M$_{\odot}$ companion
is a black hole then $a_{min} \sim$ 0.01 AU.}
\label{velocity}
\end{figure*}

\begin{figure*}
\begin{center}
\begin{tabular}{ccc}
\includegraphics[width=0.52\textwidth]{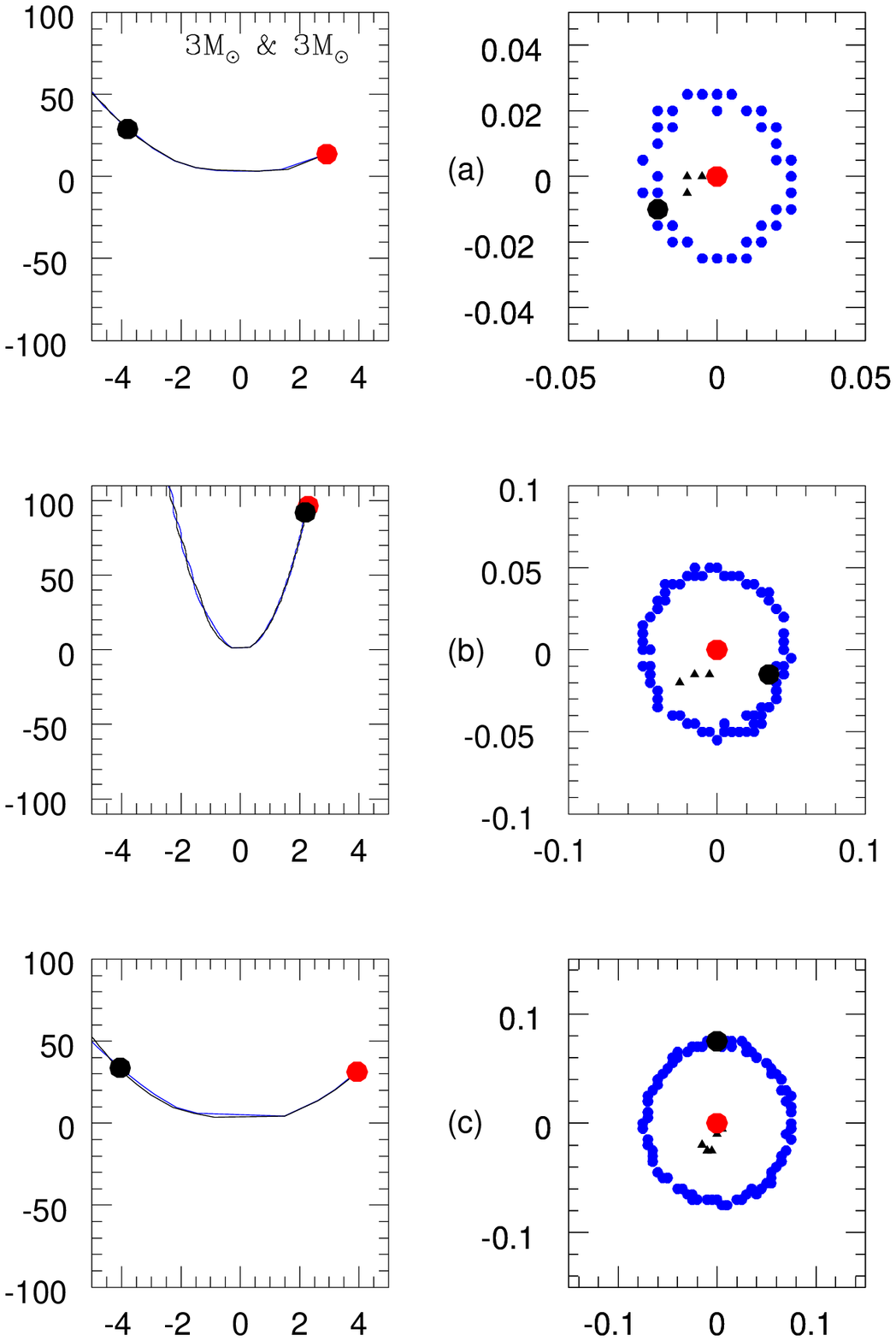}
&
\includegraphics[width=0.52\textwidth]{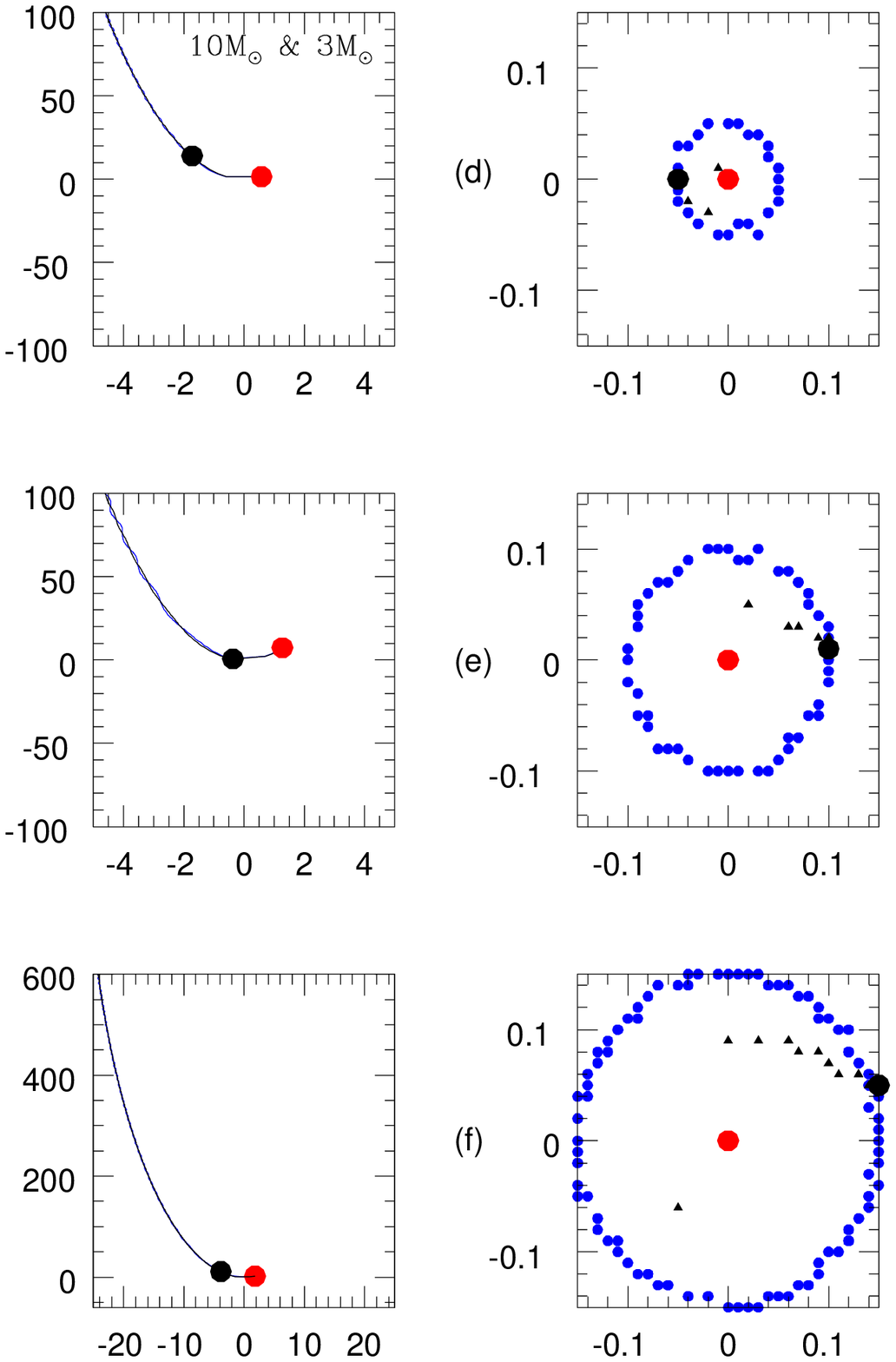}
&
\\
\end{tabular}
\end{center}
\caption{Orbits of stars (in units of AU) prior to collision for binaries
of 3M$_{\odot}$ \& 3M$_{\odot}$ in panels $(a) - (c)$ and binaries of 
10M$_{\odot}$ \& 3M$_{\odot}$ in panels $(d) - (f)$. For all panels, the 
graphs on the left are the orbits of the stars as they pass near the SMBH 
located at the origin. For panels $(a) - (c)$, the graphs on the right are 
plotted at the center of mass frame.  For panels $(d) - (f)$, the graphs 
on the right are plotted at the rest frame of the 10M$_{\odot}$ star. 
The red filled circle denotes the collision instant and the black filled
circle denotes the time when the stars started to move towards each other
as a result of the SMBH tidal force.  Note that before the binary is
disrupted, the 3M$_{\odot}$ star makes many revolutions as denoted by the
blue circles. After the binary is disrupted, the approaching stars are
denoted by black triangles for clarity.  Panels $a$ and $d$ both have an
initial separation of $a$ = 0.05 AU.  
Panels $b$ and $e$ have $a$ = 0.10 AU, and panels $c$ and $f$ have $a$ =
0.15 AU.}
\label{orbit}
\end{figure*}

Under some circumstances, the binary is disrupted in such a way that the
two stars collide. Assuming that the impulsive kick is given by the SMBH
towards a random direction within the orbital plane and ignoring
gravitational focusing (which is important at low speeds), the probability
for a collision in a case that otherwise would have produced a HVS is four
time the radius of a star divided by the circumference of a circle with a
radius equal to the binary separation.  The likelihood for a collision is
expected to be smaller in the more general case where the binary lies in a
different plane than its orbit around the SMBH, unless gravitational
focusing dominates. Table \ref{tab} summarizes the actual statistical
results from our runs.  

The two stars would merge as a result of the collision if their relative
speed is lower than the escape speed from their surface ($\sim 500~{\rm
km~s^{-1}}$).  In our runs 22\% of all collisions have impact velocities
low enough to allow the two stars to coalesce (see Table \ref{tab2}).  Also
of note is the fact that many collisions involve hypervelocities of $v >$
1000 km$s^{-1}$ upon impact.  The typical impact velocity of the two stars
can be crudely estimated from a model in which the SMBH removes the angular
momentum from the binary and causes the two stars to fall toward each other
from their initial orbital separation.  Conservation of energy
\begin{align}
E = {\frac{1}{2}}{\frac{m_1m_2}{m_1+m_2}}{\dot{r}}^2 - {\frac{Gm_1m_2}{r}}
= const ,
\label{eq:energy}
\end{align}
yields the relative velocity upon impact,
\begin{align}
v_f = \left[2G(m_1+m_2){(\frac{1}{a_{min}}}-{\frac{1}{a}})\right]^{1/2}.
\label{eq:velocity}
\end{align}
The actual impact speed would vary around this value due to the additional
velocity induced by the SMBH tidal force along the axis connecting the
stars. Nevertheless, Equation \ref{eq:velocity} agrees well with the median
of the distribution of impact speeds in our runs (see Figure
\ref{velocity}).  Collisions always occur shortly after tidal disruption,
as seen from the separation of the black and filled circles in Figure
\ref{orbit}.

\begin{table}
\begin{tabular}{|r|r|r|r|r|r|}
\hline
$a$ (AU)&P(3$M_{\odot})$&P(10$M_{\odot}$)\\
\hline
0.05&0.46&0.13\\
0.10&0.19&0.05\\
0.15&0.08&0.00\\
0.20&0.00&0.00\\
\hline
0.05&N/A&N/A\\
0.10&0.81&0.33\\
0.15&0.93&0.05\\
0.20&0.25&0.03\\
\hline
\end{tabular}
\caption{Probability of coalescence upon collision for different values of
$a$ for binaries of 3M$_{\odot}$ \& 3M$_{\odot}$ (second column) and
10M$_{\odot}$ \& 3M$_{\odot}$ (third column).  The top four rows show the
values obtained from our simulations for $a_{min}$ = 0.02 AU, and the
bottom rows for $a_{min}$ = 0.04 AU. Here we assume that the two stars will
merge if $v_{imp} \la$ 500 km~s$^{-1}$.}
\label{tab2}
\end{table}

\section{Fate of the Coalescing Binary}
Stellar collisions are likely the main assembly line of blue stragglers
(see e.g. \citealt{Leonard:89}; \citealt{Bailyn-Pinsonneault:95};
\citealt{Lombardi:02}), and ultracompact X-ray binaries
(e.g. \citealt{Ivanova:05}; \citealt{Lombardi:06}) in globular
clusters. The Galactic Centre of the Milky Way is another place where
collisions are likely to occur.  Tidal disruptions of a binary by the SMBH
will produce $\sim$ 0.1 collisions per HVS (see Paper I).  The ultimate
fate of the binary depends on the velocity of its member stars upon impact.
As evident from Figure \ref{velocity}, the impact velocity $v_{\rm imp}$
can vary over a wide range of values.  A star with $v_{imp} > v_{esc}\equiv
[2G(m_1+m_2)/(R_1+R_2)]^{1/2}$ will likely pass through the other star,
even during a head-on collision \citep{Freitag-Benz:05}.  However, in any
collision there certainly will be interactions where the smaller star may
gain mass and the larger star will likely lose mass
\citep{Freitag-Benz:05}.  Furthermore, a collision where the impact
velocity is less than the escape velocity $v_{esc}$ will not necessarily
end as a merger.  A grazing collision might result in envelope-ejection but
no core merger, whereas a head-on collision might result in core merger,
and thus form a more massive star \citep{Dale-Davies:06}.  The results of
smoothed particle hydrodynamics simulations of blue stragglers
\citep{Sills:05} show that off-axis collisions initially have large angular
momentum but eventually lose it to allow the merger to contract down to the
main sequence.  Off-axis collisions, which are more probable than
head-on collisions, could nevertheless lead to HVSs which are rapidly
spinning \citep{Alexander-Kumar:01}. Finally, a merger between a lower mass
star with a higher mass star may extend the massive star's main-sequence
phase \citep{Dale-Davies:06}.

\citet{Edelmann:05} notes that HVS HE 0437-5439 might be the merger of two
4M$_{\odot}$ stars, and that such a merger is consistent with the age of
the HVS.  Furthermore, \citet{Hirsch:05} suggests that HVS US 708, a
subluminous O star, might be the merger of two helium-core white
dwarfs. After losing mass, some of the coalescing binaries in our runs
might end up with sufficient velocity to escape the galaxy.  Unfortunately,
our simulations cannot treat mass loss as well as the large amount of
thermal energy deposited in each collision (see
\citealt{Leonard-Livio:95}).  Coalescing binary systems that remain bound
to the SMBH could end up as massive S-stars.

Aside from stellar collisions of main-sequence stars, it is possible for
collisions to involve compact objects. As long as the colliding objects are
gravitationally bound, the compact object will eventually settle to the
center of the merger remnant due to angular momentum transport by dynamical
friction (gravitationally-induced spiral arms) or viscosity in the stellar
envelope (see e.g. the simulation of a black hole-helium star merger in
\citealt{Zhang-Fryer:01}) as well as gravitational radiation.  The
situation of a stellar mass black hole surrounded by a star also appears in
the collapsar progenitors of long-duration Gamma-Ray Bursts (GRBs)
\citep{Macfadyen-Woosley:99}.  However, these events result from the
collapse of the stellar core and so the accretion rate into the black hole
is larger than the Eddington limit by more than 10 orders of magnitude
\citep{Narayan-Piran:01}.  In our case, the accretion might be limited by
the Eddington luminosity and so the resulting sources would be much fainter
than GRBs.  It is unclear whether this accretion would result in an
implosion or an explosion.  In the case of a neutron star companion, one
gets a Thorne-\.{Z}ytkow object, with a similar accretion rate
\citep{Pod:96}.

\section{Conclusions} \label{Im}

Our N-body simulations show that tight binaries with a separation between
0.05-0.20 AU which approach within a distance $\simeq$ 10 AU from SgrA*
could produce both HVSs and collisions (in $\sim 10\%$ of all cases) where
occasionally the two objects may coalesce.  Coalescence occurs if the impact
velocity is sufficiently low \citep{Freitag-Benz:05}, otherwise the SMBH
would likely eject the two stars into separate orbits (see Paper I).  The
large variance in collision velocity shown in Fig. \ref{velocity} will result 
in a broad range of collision products.  

Given the low production rate of HVSs in the Milky-Way galaxy $\sim
10^{-5}~{\rm yr}^{-1}$ (Brown et al. 2006b), and that collisions occur in
$\sim 10\%$ of all cases with mergers accounting for $\sim 20\%$ of all
collisions, the expected production rate of S-stars from binary mergers is
once every 5 million years.  Since this period is comparable to the
lifetime of the massive S-stars, most of these stars can not be the merger
remnants of binary systems.  It is unlikely that any of these collision
events will be observed in real time. Following a coalescence event, the
remnant will appear different from main-sequence stars of the same mass
only if the progenitors evolved substantially before their collision time.

\section*{Acknowledgments}

We thank Avery Broderick, Warren Brown, Yosi Gelfand, and Loren Hoffman for 
their time and useful discussions, and our referee for helpful suggestions. 
This work was supported in part by Harvard University funds.

\bibliographystyle{mn2e.bst}
\bibliography{Paper.bib}

\begin{thebibliography}{}

\bibitem[\protect\citeauthoryear{{Aarseth}}{{Aarseth}}{1999}]{Aarseth:99}
{Aarseth} S.~J.,  1999, \pasp, 111, 1333

\bibitem[\protect\citeauthoryear{{Abt}}{{Abt}}{1983}]{Abt:83}
{Abt} H.~A.,  1983, \araa, 21, 343

\bibitem[\protect\citeauthoryear{{Alexander} \& {Kumar}}{{Alexander} \&
  {Kumar}}{2001}]{Alexander-Kumar:01}
{Alexander} T.,  {Kumar} P.,  2001, \apj, 549, 948

\bibitem[\protect\citeauthoryear{{Bailyn} \& {Pinsonneault}}{{Bailyn} \&
  {Pinsonneault}}{1995}]{Bailyn-Pinsonneault:95}
{Bailyn} C.,  {Pinsonneault} M.,  1995, \apj, 439, 705

\bibitem[\protect\citeauthoryear{{Bromley}, {Kenyon}, {Geller}, {Barcikowski},
  {Brown} \& {Kurtz}}{{Bromley} et~al.}{2006}]{Bromley:06}
{Bromley} B.,  {Kenyon} S.,  {Geller} M.,  {Barcikowski} E.,  {Brown} W.,
  {Kurtz} M.,  2006, (astro-ph/0608159)

\bibitem[\protect\citeauthoryear{{Brown} W.R.and~{Geller}, {Kenyon} \&
  {Kurtz}}{{Brown} et~al.}{2006a}]{Brown:06a}
{Brown} W.R.and~{Geller} M.,  {Kenyon} S.,    {Kurtz} M.,  2006a, \apjl, 640,
  L35

\bibitem[\protect\citeauthoryear{{Brown} W.R.and~{Geller}, {Kenyon} \&
  {Kurtz}}{{Brown} et~al.}{2006b}]{Brown:06b}
{Brown} W.R.and~{Geller} M.,  {Kenyon} S.,    {Kurtz} M.,  2006b, \apj, 647,
  303

\bibitem[\protect\citeauthoryear{{Brown}, {Geller}, {Kenyon} \&
  {Kurtz}}{{Brown} et~al.}{2005}]{Brown:05}
{Brown} W.,  {Geller} M.,  {Kenyon} S.,    {Kurtz} M.,  2005, \apjl, 622, L33

\bibitem[\protect\citeauthoryear{{Dale} \& {Davies}}{{Dale} \&
  {Davies}}{2006}]{Dale-Davies:06}
{Dale} J.,  {Davies} M.,  2006, \mnras, 366, 1424

\bibitem[\protect\citeauthoryear{{Duquennoy} \& {Mayor}}{{Duquennoy} \&
  {Mayor}}{1991}]{Duquennoy-Mayor:91}
{Duquennoy} A.,  {Mayor} M.,  1991, \aap, 248, 485

\bibitem[\protect\citeauthoryear{{Eckart} \& {Genzel}}{{Eckart} \&
  {Genzel}}{1997}]{Eckart-Genzel:97}
{Eckart} A.,  {Genzel} R.,  1997, \mnras, 284, 576

\bibitem[\protect\citeauthoryear{{Edelmann}, {Napiwotzki}, {Heber},
  {Christlieb} \& {Reimers}}{{Edelmann} et~al.}{2005}]{Edelmann:05}
{Edelmann} H.,  {Napiwotzki} R.,  {Heber} U.,  {Christlieb} N.,    {Reimers}
  D.,  2005, \apjl, 634, L181

\bibitem[\protect\citeauthoryear{{Freitag} \& {Benz}}{{Freitag} \&
  {Benz}}{2005}]{Freitag-Benz:05}
{Freitag} M.,  {Benz} W.,  2005, \mnras, 358, 1133

\bibitem[\protect\citeauthoryear{{Fuentes}, {Stanek}, {Gaudi}, {McLeod},
  {Bogdanov}, {Hartman}, {Hickox} \& {Holman}}{{Fuentes}
  et~al.}{2005}]{Fuentes:05}
{Fuentes} C.,  {Stanek} K.,  {Gaudi} B.,  {McLeod} B.,  {Bogdanov} S.,
  {Hartman} J.,  {Hickox} R.,    {Holman} M.,  2005, \apjl, submitted
  (astro-ph/0507520)

\bibitem[\protect\citeauthoryear{{Ghez}, {Salim}, {Hornstein}, {Tanner}, {Lu},
  {Morris}, {Becklin} \& {Duch\^{e}ne}}{{Ghez} et~al.}{2005}]{Ghez:05}
{Ghez} A.,  {Salim} S.,  {Hornstein} S.,  {Tanner} A.,  {Lu} J.,  {Morris} M.,
  {Becklin} E.,    {Duch\^{e}ne} G.,  2005, \apj, 620, 744

\bibitem[\protect\citeauthoryear{{Ginsburg} \& {Loeb}}{{Ginsburg} \&
  {Loeb}}{2006}]{Ginsburg:06}
{Ginsburg} I.,  {Loeb} A.,  2006, \mnras, 368, 221 (Paper I)

\bibitem[\protect\citeauthoryear{{Gualandris}, {Portegies Zwart} \&
  {Sipior}}{{Gualandris} et~al.}{2005}]{Gualandris:05}
{Gualandris} A.,  {Portegies Zwart} S.,    {Sipior} M.,  2005, \mnras, 363, 223

\bibitem[\protect\citeauthoryear{{Hansen} \& {Milosavljevi{\'c}}}{{Hansen} \&
  {Milosavljevi{\'c}}}{2003}]{Hansen:03}
{Hansen} B.,  {Milosavljevi{\'c}} M.,  2003, \apjl, 593, L77

\bibitem[\protect\citeauthoryear{{Heacox}}{{Heacox}}{1998}]{Heacox:98}
{Heacox} W.,  1998, \aj, 115, 325

\bibitem[\protect\citeauthoryear{{Hills}}{{Hills}}{1988}]{Hills:88}
{Hills} J.,  1988, \nat, 331, 687

\bibitem[\protect\citeauthoryear{{Hirsch}, {Heber}, {O'Toole} \&
  {Bresolin}}{{Hirsch} et~al.}{2005}]{Hirsch:05}
{Hirsch} H.,  {Heber} U.,  {O'Toole} S.,    {Bresolin} F.,  2005, \aap, 444,
  L61

\bibitem[\protect\citeauthoryear{{Ivanova}, {Rasio}, {Lombardi}, {Dooley},  \&
  {Proulx}}{{Ivanova} et~al.}{2005}]{Ivanova:05}
{Ivanova} N.,  {Rasio} F.,  {Lombardi} J.,  {Dooley} K.,     {Proulx} Z.,
  2005, \apjl, 621, L109

\bibitem[\protect\citeauthoryear{{Lada}}{{Lada}}{2006}]{Lada:06}
{Lada} C.,  2006, \apjl, 640, L63

\bibitem[\protect\citeauthoryear{{Larson}}{{Larson}}{2003}]{Larson:03}
{Larson} R.,  2003, \rpp, 66, 1651

\bibitem[\protect\citeauthoryear{{Leonard}}{{Leonard}}{1989}]{Leonard:89}
{Leonard} P.,  1989, \aj, 98, 217L

\bibitem[\protect\citeauthoryear{{Leonard} \& {Livio}}{{Leonard} \&
  {Livio}}{1995}]{Leonard-Livio:95}
{Leonard} P.,  {Livio} M.,  1995, \apjl, 447, L121

\bibitem[\protect\citeauthoryear{{Lombardi}, {Proulx}, {Dooley}, {Theriault},
  {Ivanova} \& F.A.}{{Lombardi} et~al.}{2006}]{Lombardi:06}
{Lombardi} J.,  {Proulx} Z.,  {Dooley} K.,  {Theriault} E.,  {Ivanova} N.,
  F.A. R.,  2006, \apj, 640, 441

\bibitem[\protect\citeauthoryear{{Lombardi}, {Warren}, {Rasio}, {Sills} \&
  {Warren}}{{Lombardi} et~al.}{2002}]{Lombardi:02}
{Lombardi} J.,  {Warren} J.,  {Rasio} F.,  {Sills} A.,    {Warren} A.,  2002,
  \apj, 568, 939

\bibitem[\protect\citeauthoryear{{MacFadyen} \& {Woosley}}{{MacFadyen} \&
  {Woosley}}{1999}]{Macfadyen-Woosley:99}
{MacFadyen} A.,  {Woosley} S.,  1999, \apj, 524, 262

\bibitem[\protect\citeauthoryear{{Milosavljevi{\'c}} \&
  {Loeb}}{{Milosavljevi{\'c}} \& {Loeb}}{2004}]{M-Loeb:04}
{Milosavljevi{\'c}} M.,  {Loeb} A.,  2004, \apjl, 604, L45

\bibitem[\protect\citeauthoryear{{Narayan}, {Piran} \& {Kumar}}{{Narayan}
  et~al.}{2001}]{Narayan-Piran:01}
{Narayan} R.,  {Piran} T.,    {Kumar} P.,  2001, \apj, 557, 949

\bibitem[\protect\citeauthoryear{{Podsiadlowski}}{{Podsiadlowski}}{1996}]{Pod:%
96}
{Podsiadlowski} P.,  1996, IAUS, 165, 29

\bibitem[\protect\citeauthoryear{{Reid} \& {Brunthaler}}{{Reid} \&
  {Brunthaler}}{2004}]{Reid-Brunthaler:04}
{Reid} M.,  {Brunthaler} A.,  2004, \apj, 616, 872

\bibitem[\protect\citeauthoryear{{Sch\"{o}del}, {Ott}, {Genzel}, {Eckart},
  {Mouawad} \& {Alexander}}{{Sch\"{o}del} et~al.}{2003}]{Scho:03}
{Sch\"{o}del} R.,  {Ott} T.,  {Genzel} R.,  {Eckart} A.,  {Mouawad} N.,
  {Alexander} T.,  2003, \apj, 596, 1015

\bibitem[\protect\citeauthoryear{{Sills}, {Adams} \& {Davies}}{{Sills}
  et~al.}{2005}]{Sills:05}
{Sills} A.,  {Adams} T.,    {Davies} M.,  2005, \mnras, 358, 716

\bibitem[\protect\citeauthoryear{{Yu} \& {Tremaine}}{{Yu} \&
  {Tremaine}}{2003}]{Yu-Tremaine:03}
{Yu} Q.,  {Tremaine} S.,  2003, \apj, 599, 1129

\bibitem[\protect\citeauthoryear{{Zhang} \& {Fryer}}{{Zhang} \&
  {Fryer}}{2001}]{Zhang-Fryer:01}
{Zhang} W.,  {Fryer} C.,  2001, \apj, 550, 357

\end{thebibliography}

\bsp

\end{document}